\documentclass{ecai} 

\usepackage{latexsym}
\usepackage{amssymb}
\usepackage{amsmath}
\usepackage{amsthm}
\usepackage{booktabs}
\usepackage{enumitem}
\usepackage{graphicx}
\usepackage{color}
\usepackage{array}

\newcommand{\BibTeX}{B\kern-.05em{\sc i\kern-.025em b}\kern-.08em\TeX}

\begin{document}


\begin{frontmatter}


\paperid{123} 


\title{TAIBOM: Bringing Trustworthiness \\ to AI-Enabled Systems}


\author[A]{\fnms{Vadim}~\snm{Safronov}\orcid{0009-0005-6431-0125}\thanks{Corresponding Author. Email: vadim.safronov@cs.ox.ac.uk}}
\author[B]{\fnms{Anthony}~\snm{McCaigue}\orcid{0009-0008-8679-6578}}
\author[B]{\fnms{Nicholas}~\snm{Allott}\orcid{0000-0001-7473-0565}}
\author[A]{\fnms{Andrew}~\snm{Martin}\orcid{0000-0002-8236-980X}}


\address[A]{University of Oxford}
\address[B]{NquiringMinds}


\begin{abstract}
The growing integration of open-source software and AI-driven technologies has introduced new layers of complexity into the software supply chain, challenging existing methods for dependency management and system assurance. While Software Bills of Materials (SBOMs) have become critical for enhancing transparency and traceability, current frameworks fall short in capturing the unique characteristics of AI systems --- namely, their dynamic, data-driven nature and the loosely coupled dependencies across datasets, models, and software components. These challenges are compounded by fragmented governance structures and the lack of robust tools for ensuring integrity, trust, and compliance in AI-enabled environments. 

In this paper, we introduce Trusted AI Bill of Materials (TAIBOM) --- a novel framework extending SBOM principles to the AI domain. TAIBOM provides (i) a structured dependency model tailored for AI components, (ii) mechanisms for propagating integrity statements across heterogeneous AI pipelines, and (iii) a trust attestation process for verifying component provenance. We demonstrate how TAIBOM supports assurance, security, and compliance across AI workflows, highlighting its advantages over existing standards such as SPDX and CycloneDX. This work lays the foundation for trustworthy and verifiable AI systems through structured software transparency.
\end{abstract}

\end{frontmatter}


\section{Introduction}
\label{intro}
The rapid expansion of open-source software and AI-driven technologies has introduced unprecedented complexity into the software supply chain. Managing dependencies is already a challenging task, but the integration of AI with digital infrastructures further amplifies the difficulty~\cite{SC_motivation_1, SC_motivation_2}. In response, Software Bills of Materials (SBOMs) have emerged as a mechanism for improving accountability and traceability across software ecosystems~\cite{Biden2021CybersecuritySBOM, EU_SBOM_Work2023}.

While various organisations, researchers, and developers acknowledge the growing importance and usefulness of SBOMs, the absence of mature tools for their generation and application in assurance, security, and compliance remains a significant gap~\cite{SC_motivation_1, SBOM_needs_1, SBOM_needs_2}. This challenge is particularly significant in AI-enabled systems, where traditional SBOM frameworks struggle to accommodate AI-specific complexities. AI models are inherently dynamic and data-driven, continuously evolving through updates and retraining. Ensuring stability and version control is a persistent challenge. Furthermore, AI systems comprise multiple interdependent components — training datasets, training software, refinement datasets, trained weights, inference software — that are often loosely coupled, making dependency tracking and provenance verification difficult. The governance of AI systems is also complicated by their distributed nature, where data ownership, model training, and system deployment often reside with different entities, making the enforcement of security, privacy, and safety standards particularly challenging.

Although there is much interest in Trust in AI~\cite{Trust_AI_1, Trust_AI_2, Trust_AI_3}, much of the discussion is predicated on the accurate identification of (and so tamer-proofing of) all these diverse aforementioned elements, which in turn raises several challenging questions. Can the integrity of model producers be verified? Can datasets --- often assembled from multiple sources --- be trusted? Can transparency and accountability be ensured throughout the AI development pipeline?

To address these challenges, we introduce the Trusted AI Bill of Materials (TAIBOM) --- an evolution of SBOM frameworks extended to AI-enabled software systems. TAIBOM is accompanied by an implementation toolkit~\footnote{https://github.com/nqminds/Trusted-AI-BOM/}, designed to support its principles. The key contributions of TAIBOM are:
\begin{itemize}
    \item A structured dependency model that conceptualises AI software components and their interrelationships.
    \item A framework for propagating integrity statements across disconnected AI environments (e.g., dataset, training, and inference), ensuring a continuous chain of trust.
    \item A trust attestation mechanism to verify the provenance and trustworthiness of AI components.
\end{itemize}

This paper is organised as follows. Section~\ref{background} overviews SBOMs and motivates the need for AIBOMs. Section~\ref{related} reviews related work, identifies limitations in existing AIBOM standards regarding trust guarantees, and outlines how TAIBOM addresses them. Section~\ref{arch} presents the TAIBOM dependency model and its component relationships. Section~\ref{use_cases} demonstrates TAIBOM’s integration into AI workflows across assurance, security, compliance, and risk management use cases, and compares its functionality to state-of-the-art AIBOM approaches. Section~\ref{discussions} discusses current limitations and future directions. Section~\ref{conclusions} concludes the paper.

\section{Background}
\label{background}

To introduce the need for trust in AI-enabled systems, this background section reviews the concept of Software Bills of Materials (SBOMs), their role in traditional software supply chains, and how they inform the emerging notion of AI-specific Bills of Materials (AIBOMs).

\subsection{Software Bill of Materials}
The Software Bill of Materials (SBOM), inspired by manufacturing industry practices, was formalised in 2018 by the US National Telecommunications and Information Administration (NTIA) to enhance software security practices and has since evolved~\cite{SBOM_background_1}. An SBOM provides a detailed inventory of software components, specifying their origins, dependencies, and references to known or potential vulnerabilities. Key elements in the SBOM ecosystem include the Common Platform Enumeration (CPE, a standardised naming scheme for software components~\cite{CPE}), Common Vulnerabilities and Exposures (CVE, a repository of publicly disclosed cybersecurity vulnerabilities~\cite{CVE}), and Common Weakness Enumeration (CWE, a classification system for software weaknesses~\cite{CWE}).
 
SBOMs ensure that all software components are identified and traceable. However, concerns remain about their granularity. Various SBOM generation approaches have emerged. Binary-focused tools~\cite{B2SFinder, OSSPolice, BAT} analyse compiled binaries using embedded metadata, string literals, and language-specific features to identify dependencies. Metadata-based tools~\cite{Trivy_SBOM-tool, Microsoft_SBOM-tool, Syft_SBOM-tool} extract dependency information from package metadata, build files, and container images. Source code analysis tools~\cite{b14, tamer, centris} inspect repositories to uncover dependencies, including hidden ones associated with known CVEs.

\subsection{AI Bill of Materials}

While existing SBOM approaches are suitable for traditional software, they fall short in addressing vulnerabilities specific to AI-enabled systems. Some argue that AI is merely another category of software and can be managed under current SBOM regulations. However, recent research suggests otherwise, highlighting a growing demand for an AI Bill of Materials (AIBOM) capable of describing and tracing AI-specific dependencies~\cite{SBOM_needs_1, AIBOM_motivation_1}.

AI software comprises traditional software components alongside AI-specific artifacts such as training data, model configurations, and inference pipelines. Unlike conventional software, AI systems evolve continuously (e.g. in response to data drift and concept drift) necessitating dynamic co-versioning registries to ensure transparency and accountability. Unlike static SBOM inventories, AI Bills of Materials (AIBOMs) must support traceable, evolving records without requiring frequent regeneration.

\section{Related Work}
\label{related}

This section reviews existing AIBOM proposals and discusses how the proposed TAIBOM approach addresses their limitations in establishing verifiable trust within AI software supply chains.

\subsection{Existing AIBOM Proposals}

Several AIBOM solutions have been proposed. Model cards, such as those by Hugging Face~\cite{huggingface_modelcards} and Google~\cite{google_modelcards}, provide metadata on datasets and models but lack comprehensive provenance tracking (e.g., training data lineage or tampering history). SBOM extensions, including CycloneDX~\cite{cyclonedx_ai} and SPDX~\cite{spdx_ai}, adapt existing SBOM formats for AI, yet their trust guarantees remain limited and often unverifiable. ML-specific tools such as DVC~\cite{dvc} and MLflow~\cite{MLflow} support lineage tracking for AI models but do not enforce verifiable trust mechanisms across the full AI development pipeline.

A more detailed comparative evaluation of these approaches, including their support for provenance, tamper detection, and dependency analysis, is presented in Section~\ref{use_cases}.

\subsection{TAIBOM: A Trustable AIBOM Solution}

Existing AIBOM solutions lack trust guarantees — there are no established techniques to verify dataset provenance or ensure model integrity throughout the AI workflow. TAIBOM addresses this gap by introducing a trust-enabled AIBOM data model that represents key AI artifacts (including Data, Code, and AI System objects) along with their interconnections and trust relationships. As detailed in Section~\ref{arch}, TAIBOM integrates cryptographic attestations, integrity verification, and dataset provenance tracking to establish verifiable trust across AI-enabled software supply chains. By design, TAIBOM adopts a general and extensible structure to accommodate a wide range of AI-enabled software systems — including those incorporating frontier AI such as Large Language Models (LLMs) and other forms of generative AI.

\section{TAIBOM Architecture}
\label{arch}

The main purpose of TAIBOM is to provide a structured and verifiable framework for managing trust in AI software components, ensuring their provenance, integrity and traceability across the entire AI software lifecycle.

\begin{figure*}[t]
  \centering
  \includegraphics[width=\textwidth]{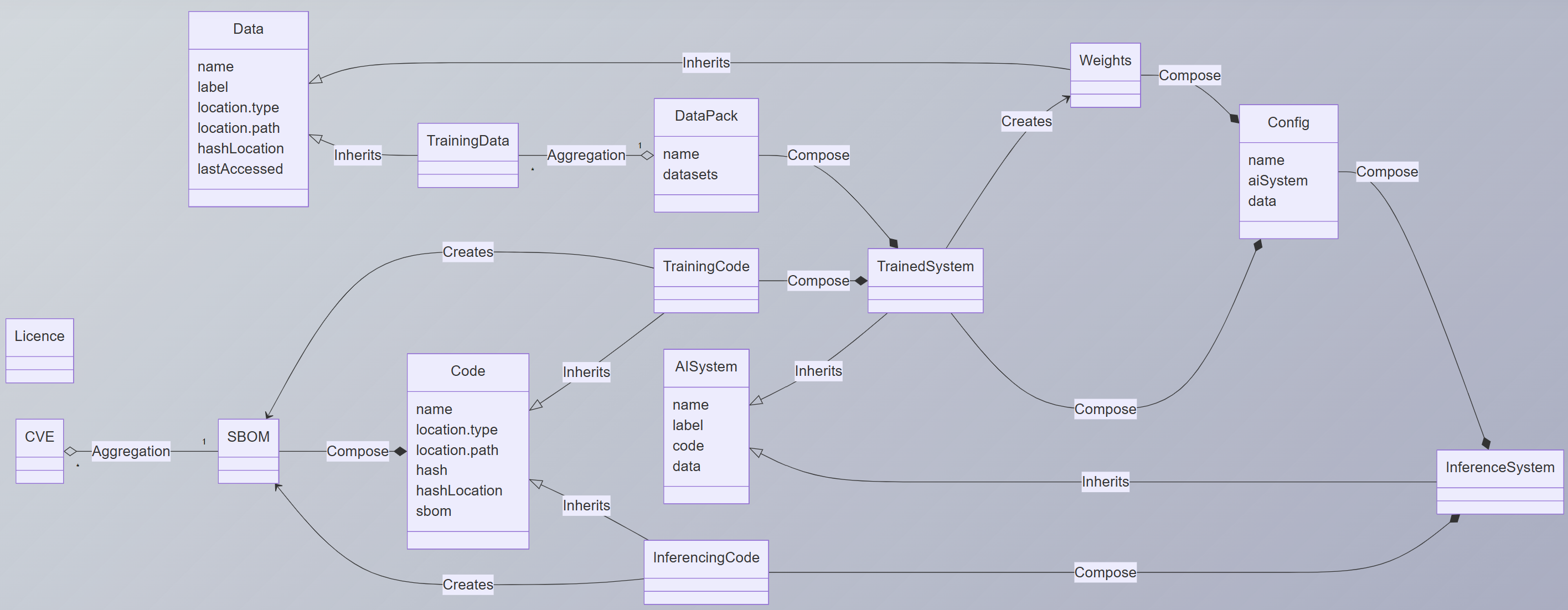} 
  \caption{TAIBOM Data Model.}
  \label{fig:interarrival_final}
\end{figure*}

\subsection{TAIBOM Data Model}

The TAIBOM data model consists of multiple interconnected class objects depicted at Figure 1.
 
\textbf{Data.} \texttt{Data} is a parent class for all datasets encompassing metadata such as name, label, location, cryptographic hashes, and last access time. \texttt{TrainingData} extends \texttt{Data}, specifically capturing AI training datasets, while \texttt{DataPack} aggregates multiple \texttt{TrainingData} instances, representing a structured collection of datasets used in AI model training.
 
\textbf{Code.} \texttt{Code} is a parent class for all code-related objects which encapsulates all software artifacts, tracking their location, cryptographic integrity, and SBOM references. \texttt{TrainingCode} and \texttt{InferencingCode} extends the parent Code component and are used to identify and describe training and inference software along with their SBOMs. \texttt{SBOM} structure further integrates with CVE identifiers, ensuring that known vulnerabilities are explicitly linked to the AI system’s components. Additionally, \texttt{License} metadata is incorporated to track software licensing requirements.
 
\textbf{AI System.} \texttt{AISystem} is a parent class that encompasses labels, code references, and training datasets. The \texttt{TrainedSystem} and \texttt{InferenceSystem} inherits from \texttt{AISystem}, serving as the components responsible for executing model training and inference respectively. The \texttt{TrainedSystem} component integrates \texttt{DataPack} and \texttt{TrainingCode}, and producing \texttt{Weights}, which represent the learned model parameters. \texttt{Weights} inherit from \texttt{Data} and are linked to \texttt{Config}, which encapsulates key AI system parameters, including associated training data and system metadata. The \texttt{InferenceSystem} is the object of the resulting AI system, composed on \texttt{Config} and \texttt{InferencingCode}, used for designated inference tasks in the actual deployment.

\subsection{TAIBOM Framework}

A fundamental aspect of the described TAIBOM Data Model is ensuring that all critical components, including training datasets, training code packages, and SBOM descriptors, are cryptographically signed and versioned. The linkage between SBOM descriptors and corresponding training code establishes a chain of trust. Once the system undergoes training with a predefined configuration, trained weights, inference code, and the SBOM descriptor for the SBOM code itself are also signed. As all components contain signed hashes and traceback from resulting object to the source object, the architecture provides integrity, traceability, and accountability, mitigating risks associated with unauthorised modification of data, model weights, code or AI system configurations.

\begin{figure*}[t]
  \centering
  \includegraphics[width=\textwidth]{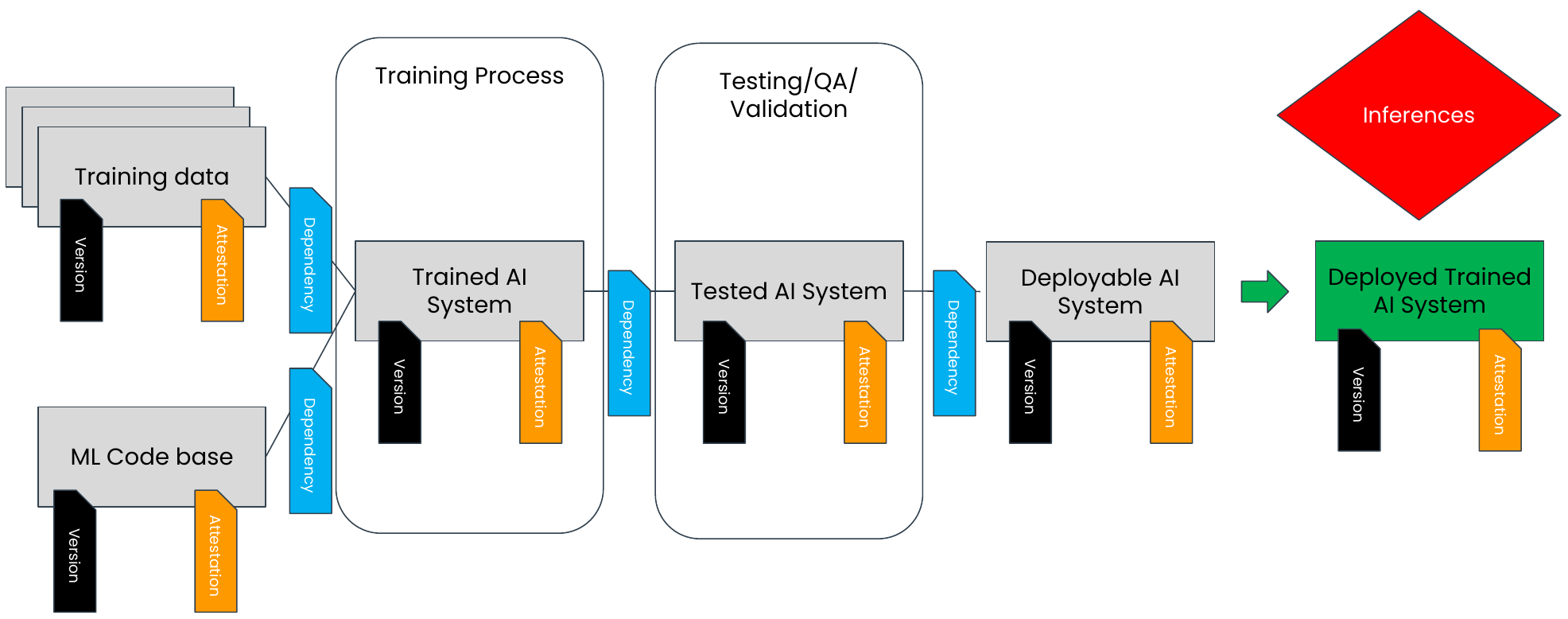} 
  \caption{TAIBOM Operation Workflow.}
  \label{fig:interarrival_final}
\end{figure*}

The TAIBOM operational workflow, illustrated in Figure 2, follows the conventional structure of machine learning system development and deployment. It is organised into several key phases, with TAIBOM augmenting each step through additional data collection, attestation, and integrity mechanisms. The training phase involves the instantiation of a trained AI system that is explicitly tied to versioned and attested training data and machine learning code, enabling full traceability of all artefacts involved. The testing, QA, and validation phase assesses the robustness of the trained system, ensuring compliance with predefined performance and security requirements. Once validated, the deployable AI system is assembled using tested and attested components to ensure stability and reproducibility. The final phase involves the deployed trained AI system, which performs real-world inferences while preserving integrity guarantees and provenance tracking as defined by the TAIBOM attestation and versioning data model.

\section{TAIBOM Use Cases}
\label{use_cases}

This section evaluates the TAIBOM framework against most widespread AI documentation and software composition tools — SPDX~\cite{spdx_ai}, CycloneDX~\cite{cyclonedx_ai} and model cards (Google~\cite{google_modelcards}, HuggingFace~\cite{huggingface_modelcards}) — by examining four critical use cases that represent operational challenge in AI lifecycle assurance, security, and traceability. The analysis contrasts the technical capabilities and design limitations of existing tools with TAIBOM’s architecture and operational semantics.

Model cards provide standardised documentation for AI models, including training configurations, intended use cases, and known limitations. However, they are self-reported and do not enforce integrity or provenance checks on datasets, training code, or derived models. CycloneDX and SPDX AI extensions offer structured software composition formats, enabling dependency visibility and limited vulnerability referencing. However, these frameworks do not establish cryptographic links between AI components (e.g., training datasets, trained weights) and their associated metadata, and thus cannot verify whether components have remained unchanged or trustworthy throughout the development lifecycle.

In contrast, TAIBOM enforces signed attestations and cryptographic integrity checks across AI artifacts, including datasets, training code, configuration files, and model weights. This ensures that all artifacts can be independently validated and traced to their origin, providing a trust dimension that is currently absent other approaches. 

Below, we detail four representative use cases (UCs) in which TAIBOM’s trust mechanisms offer substantial value. Table~\ref{tab:taibom-uc} provides a comparison of TAIBOM, Model Cards, and SPDX/CycloneDX AI extensions across some of the most common use cases in the AI model lifecycle.

\begin{table}[ht]
\scriptsize
\centering
\caption{Comparison of TAIBOM, Model Cards, and SPDX/CycloneDX for AI assurance use cases.}
\label{tab:taibom-uc}
\renewcommand{\arraystretch}{1.2}
\begin{tabular}{>{\raggedright\arraybackslash}p{1.1cm}
                >{\raggedright\arraybackslash}p{1.8cm}
                >{\raggedright\arraybackslash}p{1.8cm}
                >{\raggedright\arraybackslash}p{1.8cm}}
\toprule
\textbf{Use Case} & \textbf{TAIBOM} & \textbf{Model Cards} & \textbf{SPDX/CycloneDX} \\
\midrule
UC1: Declaring Training Data & Signed, versioned, hash-verified datasets with licence and timestamp metadata. & Descriptive only; lacks signing, versioning, or integrity verification. & Can reference datasets, but lacks formal signing or provenance tracking. \\
\midrule
UC2: Data Poisoning Detection & Verifies integrity through hash-based attestation; detects tampering. & No support for integrity checks or poisoning detection. & No dataset-level validation or tamper detection. \\
\midrule
UC3: System Tampering Detection & Runtime attestation and verification of code, model weights, and config. & No post-training verification or attestation. & Captures static dependencies only; lacks runtime validation. \\
\midrule
UC4: CVE Impact Tracing & Tracks CVEs across training data, code, and models with lifecycle propagation. & No CVE linkage or dependency tracking. & CVE mapping limited to package-level; lacks AI pipeline context. \\
\bottomrule
\end{tabular}
\end{table}


\newcommand{\code}[1]{\texttt{#1}}

\subsection*{UC 1: Declaring Training Data for Transparency}

AI model training frequently involves datasets compiled from disparate or opaque sources. Without structured, verifiable records of data composition, organisations face difficulties validating dataset provenance, understanding licensing implications, or reproducing experimental outcomes.

\subsubsection*{Existing Approaches}

\textbf{SPDX:} The SPDX standard, originally designed for traditional software, has been extended via SPDX AI to include certain AI artifacts. Datasets may be represented as generic \code{File} or \code{Package} elements using fields such as \code{name}, \code{fileName}, \code{licenseConcluded}, or \code{checksums}. However, SPDX AI treats data as static external files and lacks a domain-specific schema for dataset structure, role (e.g., training vs. test sets), or provenance tracking.

\textbf{CycloneDX:} CycloneDX AI allows datasets to be represented as components with \code{type: data}, supporting metadata like \code{externalReferences} and \code{hashes}. However, the format is flat and does not support relationships between datasets (e.g., folds or partitions), nor does it express their functional roles in training workflows.

\textbf{Model Cards:} Model cards may include narrative descriptions of datasets, their origin, and limitations, but offer no structured or verifiable representation of dataset usage.

\subsubsection*{TAIBOM Approach} 

TAIBOM defines \code{Data} and \code{TrainingData} objects, each containing cryptographic hashes, source URIs, access times, and licensing metadata. Composite datasets are grouped using the \code{DataPack} structure. These objects are cryptographically signed and versioned, and explicitly linked to the training configuration (\code{Config}) and \code{TrainedSystem} components, ensuring verifiable dataset usage.

\subsubsection*{Brief Evaluation Summary}

TAIBOM enables formal, cryptographically verifiable declarations of training datasets, while existing tools treat datasets as informal metadata or lack semantics for AI-specific data structuring.

\subsection*{UC 2: Assessing Training Data for Poisoning}

Data poisoning introduces malicious or manipulated samples into training datasets, often without detection. Detecting poisoning requires dataset versioning, reproducibility, and traceability of data inputs to model outputs.

\subsubsection*{Existing Approaches}

\textbf{SPDX:} While SPDX allows for checksums on files, these are static and not integrated with AI workflows. There is no structured support for dataset versioning or lineage tracking.

\textbf{CycloneDX:} CycloneDX supports hash-based integrity checks and can represent dataset revisions as separate components. However, there is no systematic way to express temporal lineage or bind datasets to specific models.

\textbf{Model Cards:} Model cards lack dataset version tracking or programmatic validation, and are unsuitable for forensic analysis.

\subsubsection*{TAIBOM Approach} 

Each \code{TrainingData} object includes versioning metadata and a signed cryptographic hash. TAIBOM records lineage via associations between \code{DataPack}, \code{TrainedSystem}, and \code{Weights}. Reused datasets can be detected via hash comparisons, and deviations trigger verification failures. This supports trace-based identification of potential poisoning across training runs.

\subsubsection*{Brief Evaluation Summary}

TAIBOM supports integrity verification and historical comparison of training datasets --- capabilities absent from SPDX AI, CycloneDX, and model card approaches.

\subsection*{UC 3: Detecting Training and Inference System Tampering}

Post-training tampering of AI components, such as code, configurations, or model weights, can result in erroneous or malicious behaviour. Mitigation requires strong binding between training inputs and deployed inference artifacts.

\subsubsection*{Existing Approaches}

\textbf{SPDX:} SPDX AI documents software dependencies and licenses, but does not model training or inference workflows. There is no construct to relate code to resulting models or configurations.

\textbf{CycloneDX:} CycloneDX AI introduces AI-specific components but lacks explicit workflow modeling. Dependencies may be represented via a graph, but without semantic links between artifacts (e.g., weights generated from training code).

\textbf{Model Cards:} Model cards describe configurations and limitations, but are static and decoupled from runtime code and outputs.

\subsubsection*{TAIBOM Approach}

TAIBOM introduces structured workflow representations via \code{TrainedSystem} and \code{InferenceSystem} classes. Each references \code{TrainingCode}, \code{InferencingCode}, \code{Weights}, and \code{Config}, all of which are independently signed and versioned. These links enable reproducible revalidation of deployed systems and detection of unauthorized changes to any component.

\subsubsection*{Brief Evaluation Summary}

TAIBOM supports system-level tamper detection via cryptographic linkage and provenance tracking, whereas other tools lack workflow semantics or enforcement mechanisms.

\subsection*{UC 4: Evaluating CVE Impact on Training and Inference Systems}

New CVEs may affect software libraries used in AI training or inference. Identifying affected models requires understanding which code components were used and how they relate to model artifacts.

\subsubsection*{Existing Approaches}

\textbf{SPDX:} SPDX AI allows linking to CVEs via external references, but does not relate these to model outputs or training processes. Vulnerabilities are scoped to individual files or packages only.

\textbf{CycloneDX:} CycloneDX includes a detailed vulnerability schema, enabling component-level CVE annotations. However, these annotations are not connected to AI lifecycle stages or model dependencies, limiting impact analysis.

\textbf{Model Cards:} Model cards do not include CVE information or component-level vulnerability references.

\subsubsection*{TAIBOM Approach}
TAIBOM binds each \code{TrainingCode} and \code{InferencingCode} object to a software descriptor containing CVE and CWE references. These components are directly linked to \code{TrainedSystem} and \code{InferenceSystem} objects, allowing automated identification of affected models and inference systems.

\subsubsection*{Brief Evaluation Summary}

TAIBOM enables propagation of vulnerability information across AI workflows, supporting targeted risk assessment. Other tools remain limited to surface-level vulnerability declarations without workflow integration.

\section{Discussion and Limitations}
\label{discussions}

While TAIBOM offers a structured and verifiable framework for introducing trust into AI-enabled systems, several open challenges remain for further research.

\textbf{Granularity of Component Descriptions.} As with traditional SBOMs, the level of granularity in TAIBOM can be further refined. Determining the appropriate level of detail for representing AI-specific artifacts, such as training subsets, model checkpoints, or dynamic configurations, remains an ongoing area of improvement.

\textbf{“Trusted” does not necessarily mean “secure”.} It is important to clarify that “trusted” in the context of TAIBOM does not imply complete or absolute security. Rather, it reflects adherence to a defined trust model that includes cryptographic attestations and provenance guarantees under certain assumptions --- such as the trustworthiness of dataset providers, signing authorities, and the integrity of the signing infrastructure.

\textbf{Trust Chain Recall and Recovery.} TAIBOM assumes that components in the supply chain are verifiable. However, if a component is later found to be compromised (e.g. a poisoned dataset or a tampered training script), procedures must be in place to recall or revoke the corresponding trust attestations. Future work will explore mechanisms for restructuring or rebuilding the trust chain in such scenarios, similar to revocation in certificate-based systems.

\textbf{Scalability in Signing Large-Scale Data.} Cryptographic signing of large-scale datasets, potentially petabytes in size, presents practical challenges. To address this, probabilistic or representative sampling methods (e.g., signing hashed subsets of data blocks) can be employed, enabling efficient integrity verification without the need to sign entire datasets. However, the use of stronger trust anchors remains an open consideration, particularly for deployments requiring higher assurance. Balancing the integration of such mechanisms with performance, scalability, and operational constraints is essential for practical adoption, especially in power-constrained environments.

\section{Conclusions}
\label{conclusions}

This paper introduced TAIBOM, a trust-enabled AI Bill of Materials framework that extends conventional SBOM principles with AI-specific constructs. TAIBOM offers a structured data model, cryptographic attestation mechanisms, and a provenance-aware architecture to support trust, transparency, and integrity across the AI pipeline --- from training data and model weights to deployed inference systems.

Through detailed use cases and a comparative evaluation against existing solutions such as Model Cards, SPDX AI, and CycloneDX AI, we demonstrated TAIBOM’s capacity to address emerging challenges in AI assurance, security, and compliance. By supporting both technical verification and auditability, TAIBOM enables principled trust-building in the next generation of AI-enabled software systems.

In addition to AI-specific provenance and integrity guarantees, TAIBOM’s design can also help detect and mitigate broader system-level issues, such as data corruption during transmission or storage, and reduce the impact of human errors in artifact handling.

Future work will focus on advancing the scalability of attestation mechanisms, refining trust recovery strategies, and integrating TAIBOM with industry-standard development pipelines to support broader adoption.

\section*{Acknowledgements}
TAIBOM is supported by InnovateUK under grant 10092977: TAIBOM – Trusted AI Bill of Materials.

\bibliography{references}

\begin{thebibliography}{29}
\providecommand{\natexlab}[1]{#1}
\providecommand{\url}[1]{\texttt{#1}}
\expandafter\ifx\csname urlstyle\endcsname\relax
  \providecommand{\doi}[1]{doi: #1}\else
  \providecommand{\doi}{doi: \begingroup \urlstyle{rm}\Url}\fi

\bibitem[MLf(2019)]{MLflow}
Accelerating the machine learning lifecycle with {MLflow}.
\newblock Santa Clara, CA, May 2019. USENIX Association.

\bibitem[Tri(2024)]{Trivy_SBOM-tool}
Trivy: Open source vulnerability scanner, 2024.
\newblock URL \url{https://trivy.dev/v0.33/}.
\newblock Accessed: 2024-11-28.

\bibitem[{Anchore}(n.d.)]{Syft_SBOM-tool}
{Anchore}.
\newblock Syft: A cli tool and library for generating sboms from container images and filesystems, n.d.
\newblock URL \url{https://github.com/anchore/syft}.
\newblock Accessed: 2024-11-28.

\bibitem[Biden(2021)]{Biden2021CybersecuritySBOM}
J.~Biden.
\newblock Executive order on improving the nation's cybersecurity.
\newblock \url{https://www.whitehouse.gov/briefing-room/presidential-actions/2021/05/12/executive-order-on-improving-the-nations-cybersecurity/}, 2021.
\newblock Accessed: 2024-06-20.

\bibitem[Choung et~al.(2023)Choung, David, and and]{Trust_AI_1}
H.~Choung, P.~David, and A.~R. and.
\newblock Trust in ai and its role in the acceptance of ai technologies.
\newblock \emph{International Journal of Human–Computer Interaction}, 39\penalty0 (9):\penalty0 1727--1739, 2023.
\newblock \doi{10.1080/10447318.2022.2050543}.
\newblock URL \url{https://doi.org/10.1080/10447318.2022.2050543}.

\bibitem[Corporation(2024)]{CWE}
M.~Corporation.
\newblock About common weakness enumeration, 2024.
\newblock URL \url{https://cwe.mitre.org/about/index.html}.
\newblock [Accessed: Jun. 19, 2024].

\bibitem[CycloneDX(2024)]{cyclonedx_ai}
CycloneDX.
\newblock Cyclonedx sbom standard.
\newblock \url{https://cyclonedx.org}, 2024.
\newblock Accessed: 2025-03-31.

\bibitem[Duan et~al.(2017)Duan, Bijlani, Xu, Kim, and Lee]{OSSPolice}
R.~Duan, A.~Bijlani, M.~Xu, T.~Kim, and W.~Lee.
\newblock Identifying open-source license violation and 1-day security risk at large scale.
\newblock In \emph{Proceedings of the 2017 ACM SIGSAC Conference on Computer and Communications Security}, CCS '17, page 2169–2185, New York, NY, USA, 2017. Association for Computing Machinery.
\newblock ISBN 9781450349468.
\newblock \doi{10.1145/3133956.3134048}.
\newblock URL \url{https://doi.org/10.1145/3133956.3134048}.

\bibitem[DVC(2020)]{dvc}
DVC.
\newblock Data version control.
\newblock \url{https://dvc.org}, 2020.
\newblock Accessed: 2025-03-31.

\bibitem[European~Commission(2023)]{EU_SBOM_Work2023}
D.~C. European~Commission.
\newblock Cyber resilience act, 2023.
\newblock URL \url{https://www.cisa.gov/sites/default/files/2023-09/EU%20Commission%20SBOM%20Work_508c.pdf}.
\newblock Accessed: 2024-08-27.

\bibitem[Face(2020)]{huggingface_modelcards}
H.~Face.
\newblock Model cards.
\newblock \url{https://huggingface.co/docs/hub/model-cards}, 2020.
\newblock Accessed: 2025-03-31.

\bibitem[Feng et~al.(2020)Feng, Yuan, Li, Ban, Xiao, Wang, Tang, Su, Yu, Xu, Piao, Xue, and Huo]{B2SFinder}
M.~Feng, Z.~Yuan, F.~Li, G.~Ban, Y.~Xiao, S.~Wang, Q.~Tang, H.~Su, C.~Yu, J.~Xu, A.~Piao, J.~Xue, and W.~Huo.
\newblock B2sfinder: detecting open-source software reuse in cots software.
\newblock In \emph{Proceedings of the 34th IEEE/ACM International Conference on Automated Software Engineering}, ASE '19, page 1038–1049. IEEE Press, 2020.
\newblock ISBN 9781728125084.
\newblock \doi{10.1109/ASE.2019.00100}.
\newblock URL \url{https://doi.org/10.1109/ASE.2019.00100}.

\bibitem[Gille et~al.(2020)Gille, Jobin, and Ienca]{Trust_AI_3}
F.~Gille, A.~Jobin, and M.~Ienca.
\newblock What we talk about when we talk about trust: Theory of trust for ai in healthcare.
\newblock \emph{Intelligence-Based Medicine}, 1-2:\penalty0 100001, 2020.
\newblock ISSN 2666-5212.
\newblock \doi{https://doi.org/10.1016/j.ibmed.2020.100001}.
\newblock URL \url{https://www.sciencedirect.com/science/article/pii/S2666521220300016}.

\bibitem[Hemel et~al.(2011)Hemel, Kalleberg, Vermaas, and Dolstra]{BAT}
A.~Hemel, K.~T. Kalleberg, R.~Vermaas, and E.~Dolstra.
\newblock Finding software license violations through binary code clone detection.
\newblock In \emph{Proceedings of the 8th Working Conference on Mining Software Repositories}, MSR '11, page 63–72, New York, NY, USA, 2011. Association for Computing Machinery.
\newblock ISBN 9781450305747.
\newblock \doi{10.1145/1985441.1985453}.
\newblock URL \url{https://doi.org/10.1145/1985441.1985453}.

\bibitem[Hu et~al.(2023)Hu, Xu, Fang, Wu, Yuan, Zou, and Jin]{tamer}
T.~Hu, Z.~Xu, Y.~Fang, Y.~Wu, B.~Yuan, D.~Zou, and H.~Jin.
\newblock Fine-grained code clone detection with block-based splitting of abstract syntax tree.
\newblock In \emph{Proceedings of the 32nd ACM SIGSOFT International Symposium on Software Testing and Analysis}, ISSTA 2023, page 89–100, New York, NY, USA, 2023. Association for Computing Machinery.
\newblock ISBN 9798400702211.
\newblock \doi{10.1145/3597926.3598040}.
\newblock URL \url{https://doi.org/10.1145/3597926.3598040}.

\bibitem[Jiang et~al.(2022)Jiang, Synovic, Sethi, Indarapu, Hyatt, Schorlemmer, Thiruvathukal, and Davis]{SC_motivation_2}
W.~Jiang, N.~Synovic, R.~Sethi, A.~Indarapu, M.~Hyatt, T.~R. Schorlemmer, G.~K. Thiruvathukal, and J.~C. Davis.
\newblock An empirical study of artifacts and security risks in the pre-trained model supply chain.
\newblock In \emph{Proceedings of the 2022 ACM Workshop on Software Supply Chain Offensive Research and Ecosystem Defenses}, SCORED'22, page 105–114, New York, NY, USA, 2022. Association for Computing Machinery.
\newblock ISBN 9781450398855.
\newblock \doi{10.1145/3560835.3564547}.
\newblock URL \url{https://doi.org/10.1145/3560835.3564547}.

\bibitem[Lu et~al.(2022)Lu, Zhu, Xu, Whittle, and Xing]{AIBOM_motivation_1}
Q.~Lu, L.~Zhu, X.~Xu, J.~Whittle, and Z.~Xing.
\newblock Towards a roadmap on software engineering for responsible ai.
\newblock In \emph{Proceedings of the 1st International Conference on AI Engineering: Software Engineering for AI}, CAIN '22, page 101–112, New York, NY, USA, 2022. Association for Computing Machinery.
\newblock ISBN 9781450392754.
\newblock \doi{10.1145/3522664.3528607}.
\newblock URL \url{https://doi.org/10.1145/3522664.3528607}.

\bibitem[{Microsoft}(n.d.)]{Microsoft_SBOM-tool}
{Microsoft}.
\newblock Sbom tool: Generate software bill of materials (sboms), n.d.
\newblock URL \url{https://github.com/microsoft/sbom-tool}.
\newblock Accessed: 2024-11-28.

\bibitem[Mitchell et~al.(2019)Mitchell, Wu, Zaldivar, Barnes, Vasserman, Hutchinson, Spitzer, Raji, and Gebru]{google_modelcards}
M.~Mitchell, S.~Wu, A.~Zaldivar, P.~Barnes, L.~Vasserman, B.~Hutchinson, E.~Spitzer, I.~D. Raji, and T.~Gebru.
\newblock Model cards for model reporting.
\newblock In \emph{Proceedings of the Conference on Fairness, Accountability, and Transparency (FAT* 2019)}, pages 220--229. ACM, 2019.
\newblock \doi{10.1145/3287560.3287596}.

\bibitem[NIST(2024{\natexlab{a}})]{CPE}
NIST.
\newblock {Common Platform Enumeration (CPE)}, 2024{\natexlab{a}}.
\newblock URL \url{https://nvd.nist.gov/products/cpe}.
\newblock [Accessed: Jun. 19, 2024].

\bibitem[NIST(2024{\natexlab{b}})]{CVE}
NIST.
\newblock Nist's cve process, 2024{\natexlab{b}}.
\newblock URL \url{https://nvd.nist.gov/general/cve-process}.
\newblock [Accessed: Jun. 19, 2024].

\bibitem[Omrani et~al.(2022)Omrani, Rivieccio, Fiore, Schiavone, and Agreda]{Trust_AI_2}
N.~Omrani, G.~Rivieccio, U.~Fiore, F.~Schiavone, and S.~G. Agreda.
\newblock To trust or not to trust? an assessment of trust in ai-based systems: Concerns, ethics and contexts.
\newblock \emph{Technological Forecasting and Social Change}, 181:\penalty0 121763, 2022.
\newblock ISSN 0040-1625.
\newblock \doi{https://doi.org/10.1016/j.techfore.2022.121763}.
\newblock URL \url{https://www.sciencedirect.com/science/article/pii/S0040162522002888}.

\bibitem[Project(2023)]{spdx_ai}
S.~Project.
\newblock Spdx ai: Extending the software package data exchange for ai systems.
\newblock \url{https://spdx.dev/specifications/ai/}, 2023.
\newblock Accessed: 2025-03-31.

\bibitem[Stalnaker et~al.(2024)Stalnaker, Wintersgill, Chaparro, Di~Penta, German, and Poshyvanyk]{SC_motivation_1}
T.~Stalnaker, N.~Wintersgill, O.~Chaparro, M.~Di~Penta, D.~M. German, and D.~Poshyvanyk.
\newblock Boms away! inside the minds of stakeholders: A comprehensive study of bills of materials for software systems.
\newblock In \emph{Proceedings of the IEEE/ACM 46th International Conference on Software Engineering}, ICSE '24, New York, NY, USA, 2024. Association for Computing Machinery.
\newblock ISBN 9798400702174.
\newblock \doi{10.1145/3597503.3623347}.
\newblock URL \url{https://doi.org/10.1145/3597503.3623347}.

\bibitem[Tang et~al.(2022)Tang, Xu, Liu, Wu, Yang, Li, Luo, and Liu]{b14}
W.~Tang, Z.~Xu, C.~Liu, J.~Wu, S.~Yang, Y.~Li, P.~Luo, and Y.~Liu.
\newblock Towards understanding third-party library dependency in c/c++ ecosystem.
\newblock Oct. 2022.
\newblock URL \url{https://doi.org/10.1145/3551349.3560432}.

\bibitem[Telecommunications and Administration(2021)]{SBOM_background_1}
N.~Telecommunications and I.~Administration.
\newblock Software bill of materials (sbom), 2021.
\newblock URL \url{https://www.ntia.gov/page/software-bill-materials}.
\newblock Accessed: 2024-08-27.

\bibitem[Woo et~al.(2021)Woo, Park, Kim, Lee, and Oh]{centris}
S.~Woo, S.~Park, S.~Kim, H.~Lee, and H.~Oh.
\newblock Centris: A precise and scalable approach for identifying modified open-source software reuse.
\newblock In \emph{Proceedings of the 43rd International Conference on Software Engineering}, ICSE '21, page 860–872. IEEE Press, 2021.
\newblock ISBN 9781450390859.
\newblock \doi{10.1109/ICSE43902.2021.00083}.
\newblock URL \url{https://doi.org/10.1109/ICSE43902.2021.00083}.

\bibitem[Xia et~al.(2023)Xia, Bi, Xing, Lu, and Zhu]{SBOM_needs_1}
B.~Xia, T.~Bi, Z.~Xing, Q.~Lu, and L.~Zhu.
\newblock An empirical study on software bill of materials: Where we stand and the road ahead, 2023.
\newblock URL \url{https://arxiv.org/abs/2301.05362}.

\bibitem[Zahan et~al.(2023)Zahan, Lin, Tamanna, Enck, and Williams]{SBOM_needs_2}
N.~Zahan, E.~Lin, M.~Tamanna, W.~Enck, and L.~Williams.
\newblock Software bills of materials are required. are we there yet?
\newblock \emph{IEEE Security \& Privacy}, 21\penalty0 (2):\penalty0 82--88, 2023.
\newblock \doi{10.1109/MSEC.2023.3237100}.

\end{thebibliography}

\end{document}